\documentclass[10pt,prb,twocolumn,notitlepage,showpacs]{revtex4-1}
\usepackage{amsthm,amsmath,amsfonts,amssymb,verbatim, color}
\usepackage{graphicx}
\usepackage{subfigure}
\usepackage{epsfig,slashed}
\usepackage{color}
\usepackage[colorlinks=true,citecolor=blue,
linkcolor=blue,urlcolor=blue]{hyperref}
\begin{document}

\newcommand{\Tr}{\mathop{\mathrm{Tr}}}
\newcommand{\bsigma}{\boldsymbol{\sigma}}
\renewcommand{\Re}{\mathop{\mathrm{Re}}}
\renewcommand{\Im}{\mathop{\mathrm{Im}}}
\renewcommand{\b}[1]{\mathbf{#1}}
\newcommand{\diag}{\mathrm{diag}}
\newcommand{\sign}{\mathrm{sign}}
\newcommand{\sgn}{\mathop{\mathrm{sgn}}}

\title{Superconductivity of disordered Dirac fermions in graphene}

\author{Ionut-Dragos Potirniche,$^1$ Joseph Maciejko,$^{2,3}$ Rahul Nandkishore,$^2$  and S. L. Sondhi$^{1}$}

\affiliation{$^1$Department of Physics, Princeton University, Princeton, New Jersey 08544, USA
\\$^2$Princeton Center for Theoretical Science, Princeton University, Princeton, New Jersey 08544, USA
\\$^3$Department of Physics, University of Alberta, Edmonton, Alberta, Canada T6G 2E1}

\date\today

\begin{abstract}
We numerically study the interplay between superconductivity and disorder on the graphene honeycomb lattice with on-site Hubbard attractive interactions $U$ using a spatially inhomogeneous self-consistent Bogoliubov-de Gennes (BdG) approach. In the absence of disorder there are two phases at charge neutrality. Below a critical value $U_{c}$ for attractive interactions there is a Dirac semimetal phase and above it there is a superconducting phase. We add scalar potential disorder to the system, while remaining at charge neutrality on average. Numerical solution of the BdG equations suggests that while in the strong attraction regime ($U> U_c$) disorder has the usual effect of suppressing superconductivity, in the weak attraction regime ($U < U_c$) weak disorder enhances superconductivity. In the weak attraction regime, disorder that is too strong eventually suppresses superconductivity, i.e., there is an optimal disorder strength that maximizes the critical temperature $T_c$. Our numerical results also suggest that in the weakly disordered regime, mesoscopic inhomogeneities enhance superconductivity significantly more than what is predicted by a spatially uniform mean-field theory \`a la Abrikosov-Gorkov. In this regime, superconductivity consists of rare phase-coherent superconducting islands. We also study the enhancement of the superconducting proximity effect by disorder and mesoscopic inhomogeneities, and obtain typical spatial plots of the tunneling density of states and the superfluid susceptibility that can be directly compared to scanning tunneling miscroscopy (STM) experiments on proximity-induced superconductivity in graphene.
\end{abstract}


\maketitle

\section{INTRODUCTION}
\label{sec:intro}

Quantum many-body phenomena in disordered electronic systems have fascinated condensed matter physicists for decades.\cite{Lee} An important subclass of such problems involves the nature of superconductivity in disordered materials.\cite{Abrikosov-Gorkov} Disorder is generally believed to be always detrimental to superconductivity, although $s$-wave superconductors are largely protected against it, because of the Anderson theorem.\cite{Anderson} More recent work has examined the interplay between disorder and superconductivity more carefully. The work in \cite{ghosal_inhomogeneous_2001, ghosal_role_1998, bouadim_2011} has examined s-wave
superconductivity in the negative-U Hubbard model and showed that at large
disorder superconductivity is highly inhomogeneous with the striking 
consequence that its destruction does not lead to a closing of the single particle gap. In related developments \cite{Spivak, Oreto, Lamacraft} it has been shown that inhomogeneities/mesoscopic fluctuations can mitigate the suppression of superconductivity by disorder, with the effects being particularly striking for unconventional (non-$s$-wave) superconductivity.
Finally, it has been argued in \cite{Feigelman, BumikovGornyiMirlin, dellanna} that disorder can enhance superconductivity in the vicinity of an underlying metal-insulator transition due to the multifractality of the electronic wavefunctions.

The recent discovery of two-dimensional (2D) Dirac semimetals such as graphene\cite{Geim} has provided a new platform for investigating the interplay of disorder and superconductivity. In the clean system, at charge neutrality, the onset of superconductivity is a quantum critical phenomenon, occurring only above some threshold interaction strength, with the phase transition between the Dirac semimetal and the superconductor being governed by an interesting effective field theory displaying emergent supersymmetry.\cite{grover1,grover2, ponte} It has also been pointed out\cite{nandkishore_superconductivity_2013} that for subcritical attractive interactions, disorder has the counterintuitive effect of {\it enhancing} superconductivity, such that in the presence of weak disorder the system superconducts for arbitrarily weak attraction. It was further argued that the disorder enhancement of superconductivity in disordered Dirac fermion systems should be greatly magnified by mesoscopic fluctuation effects over and above the predictions of a mean-field theory where the superconducting order parameter is assumed to be homogeneous. The interplay of disorder and interactions in Dirac fermion systems has also been examined theoretically in complementary work \cite{FosterYuzbashyan, FosterXieChou} examining the robustness of the surface states of topological superconductors. 

In this paper, we study numerically the interplay of disorder and superconductivity in a 2D Dirac fermion system --- graphene. We find evidence indicating that weak disorder enhances superconductivity when attractive interactions are weak, but suppresses superconductivity when attractive interactions are strong. Meanwhile, strong disorder always suppresses superconductivity, such that for weak attraction there is an optimal disorder strength that maximizes the critical temperature $T_c$. We derive a phase diagram in the plane of disorder strength and attraction strength, and establish that the disorder enhancement of superconductivity for weak attraction is dominated by mesoscopic fluctuations: the superconducting phase in this regime can be thought of as rare phase-coherent superconducting islands. Finally, we investigate numerically the proximity effect on disordered graphene, and produce typical spatial plots of the superfluid density and tunneling density of states that allow this picture to be directly compared with scanning tunneling microscopy (STM) experiments on graphene.

The paper is structured as follows. In Sec.~\ref{sec:Model} we present the theoretical model, which is based on the self-consistent Bogoliubov-de Gennes (BdG) formalism for a spatially inhomogeneous pairing amplitude. This model allows us to study the role of mesoscopic fluctuations in the robustness of the emergent $s$-wave superconductivity. In Sec.~\ref{sec:Clean}  we identify numerically the critical coupling $U_c$ that separates the Dirac semimetal and the superconductor in the clean system. We then obtain the full phase diagram in the plane of disorder $V$ and interaction $U$ which summarizes the interplay between superconductivity, disorder, and finite-size effects. For weak attractive interactions, the superconducting phase is not visible in the numerical results due to finite-size effects (the coherence length grows exponentially as $U$ decreases). Thus, we are restricted to a window of couplings near $U_c$. We observe that adding disorder allows superconductivity to develop even for $U < U_c$, although because of finite-size effects the disorder strength must exceed some threshold value. Meanwhile, strong disorder suppresses superconductivity, such that there is an optimal disorder strength which maximizes $T_c$ for $U< U_c$. For $U> U_c$ disorder always suppresses superconductivity. In Sec.~\ref{sec:Inhomogeneity}, we study the spatial structure of the disorder-enabled superconducting phase for $U<U_c$. Specifically, we show numerically that superconductivity is far stronger than would be predicted based on a theory that assumes the superconductivity to be homogeneous. Examining the spectral gap and the local density of states as a function of position reveals that in this regime superconductivity is highly inhomogeneous, being dominated by rare regions with strong pairing. These spatial plots may also be directly compared with STM experiments. Finally, in Sec.~\ref{sec:Proximity} we discuss the superconducting proximity effect in dirty graphene. We present numerical evidence for the enhancement of the superfluid susceptibility by weak disorder, as well as the suppression of the superfluid susceptibility for very strong disorder.

\section{Theoretical model}
\label{sec:Model}

We consider spinful electrons hopping on a 2D honeycomb lattice near half filling, with a random scalar potential and an attractive on-site interaction. The Hamiltonian is $H = H_0 + H_\textrm{int}$, where
\begin{eqnarray}\label{Hamiltonian}
H_{0} &=& -t\sum_{\langle ij\rangle,\sigma} (c_{i\sigma}^{\dagger} c_{j\sigma} + \mathrm{H.c.}) + \sum_{i,\sigma}(V_{i} - \mu)n_{i\sigma},\nonumber \\
H_{\mathrm{int}} &=& - U \sum_{i} n_{i\uparrow} n_{i\downarrow}.
\end{eqnarray}
Here $c_{i\sigma}^{\dagger}$ creates an electron with spin $\sigma$ at site $\mathbf{r}_{i}$, $t$ is the nearest-neighbor hopping matrix element, $U>0$ is the attractive (pairing) interaction, $n_{i\sigma}=c_{i\sigma}^\dag c_{i\sigma}$ is the number of electrons of spin $\sigma$ located at site $\mathbf{r}_{i}$, $\mu$ is the chemical potential, and $V_{i}$ is a random scalar potential at site $\mathbf{r}_{i}$, which is sampled from a uniform distribution $[-V,V]$ where $V$ is the disorder strength. We measure energies in units of $t$, which is equivalent to setting $t=1$.  We have numerically studied the Hamiltonian~(\ref{Hamiltonian}) for a lattice of $N=900$ sites with periodic boundary conditions. Some tests were made for lattices of up to $N=1600$ sites. For every realization of disorder, the chemical potential was chosen to keep the system at charge neutrality on average, i.e., to ensure that $\left<n\right>\equiv\sum_{i,\sigma}\left<n_{i\sigma}\right>/N = 1$.

This simple model captures three important pieces of physics. At $V=0$, it describes the clean system which is a Dirac semimetal at low interaction strengths $U$ and an $s$-wave superconductor for higher attractive interactions. At $U=0$, it reduces to the Anderson localization problem on a honeycomb lattice. For $U\neq 0$ and $V \neq 0$, which is the focus of this paper, it captures the interplay between disorder and superconductivity for Dirac fermions.

 We note the strong similarities between our model and that of Ref.~\onlinecite{ghosal_role_1998, nandkishore_superconductivity_2013}. However, Ref.~\onlinecite{ghosal_role_1998} worked with a nearest-neighbor model on a square lattice with a conventional ``parabolic'' dispersion $\propto\cos k_x+\cos k_y$, while the Dirac nature of the electrons on the half-filled honeycomb lattice will be essential to our analysis. Meanwhile, Ref.~\onlinecite{nandkishore_superconductivity_2013} studied theoretically the interplay between superconductivity and disorder for a single species of massless, spinful Dirac fermions in 2D with attractive interactions. If parity and time-reversal symmetries are to be preserved, an odd number of species of Dirac fermions can be realized on the surface of a 3D topological insulator, but not on a purely 2D lattice.\cite{nielsen1981a,nielsen1981b,niemi1983,redlich1984} Unlike Ref.~\onlinecite{nandkishore_superconductivity_2013}, we work with a model that has an even number of species of Dirac fermions, namely two species of Dirac fermions per spin.

We investigate the interplay of disorder and superconductivity within the self-consistent BdG formalism\citep{bdg_1966, ghosal_inhomogeneous_2001} that we first briefly review. In a mean-field approximation, the interaction term $H_{\mathrm{int}}$ in Eq.~(\ref{Hamiltonian}) can be decoupled in two ways, by acquiring a local density $\left< n_{i\sigma} \right> = \left< c_{i\sigma}^{\dagger} c_{i\sigma} \right>$ or a pairing amplitude $\Delta (\mathbf{r}_{i}) = -U \left<  c_{i\downarrow} c_{i\uparrow}\right>$. Because the random scalar potential $V_i$ breaks the lattice translation symmetry at the level of the Hamiltonian, we allow the pairing amplitude to be inhomogeneous. The mean-field factorization of the interaction term yields a quadratic Hamiltonian,
\begin{align}\label{HMF}
H_\textrm{MF} &= -\sum_{\left<ij\right>, \sigma} (c_{i\sigma}^{\dagger} c_{j\sigma} + c_{j\sigma}^\dagger c_{i\sigma}) + \sum_{i,\sigma}(V_{i} - \tilde{\mu}_{i})n_{i\sigma} \nonumber\\
&+ \sum_{i} \left(\Delta(\mathbf{r}_{i}) c_{i\uparrow}^{\dagger} c_{i\downarrow}^{\dagger} + \Delta^{*} (\mathbf{r}_{i}) c_{i\downarrow} c_{i\uparrow}\right),
\end{align}
where the Hartree shift in the chemical potential is accounted for by defining $ \tilde{\mu}_{i} = \mu + U \frac{\left< n_{i} \right>}{2}$, with $\left< n_{i} \right>=\sum_{\sigma} \left< n_{i\sigma} \right>$. This Hamiltonian is diagonalized by the Bogoliubov operators $\gamma_{n\sigma}$, which are defined by
\begin{eqnarray}
c_{i\uparrow} &=& \sum_{n}\left(\gamma_{n\uparrow}u_{n}(\mathbf{r}_{i}) - \gamma_{n\downarrow}^{\dagger}v_{n}^{*}(\mathbf{r}_{i})\right),\\
c_{i\downarrow} &=& \sum_{n}\left(\gamma_{n\downarrow}u_{n}(\mathbf{r}_{i}) + \gamma_{n\uparrow}^{\dagger}v_{n}^{*}(\mathbf{r}_{i})\right).
\end{eqnarray}
The coefficients $u_{n}(\mathbf{r}_{i})$ and $v_{n}(\mathbf{r}_{i})$ satisfy the normalization condition $\sum_{n} | v_{n}(\mathbf{r}_{i})|^{2} + | u_{n}(\mathbf{r}_{i})|^{2} = 1$ for each site $\b{r}_i$. The diagonalized Hamiltonian is written as $H_\textrm{MF} = \sum_{n,\sigma} \epsilon_{n} \gamma_{n\sigma}^{\dagger} \gamma_{n\sigma}$ with $\epsilon_{n} \geq 0$, the coefficients $u_{n}(\mathbf{r}_{i})$ and $v_{n}(\mathbf{r}_{i})$ are solutions of the BdG equations
\begin{eqnarray}\label{BdGeqn}
\begin{bmatrix}
\hat{H}_{K} & \hat{\Delta}  \\
\hat{\Delta}^{*} & -\hat{H}_{K}^{*}\end{bmatrix} \left[ \begin{array}{c} u_{n}(\mathbf{r}_{i}) \\ v_{n}(\mathbf{r}_{i}) \end{array} \right] = \epsilon_{n} \left[ \begin{array}{c} u_{n}(\mathbf{r}_{i}) \\ v_{n}(\mathbf{r}_{i})\end{array}\right],\end{eqnarray}
where $\hat{H}_{K} u_{n}(\mathbf{r}_{i}) = -t\sum_{\hat{a}} u_{n}(\mathbf{r}_{i}+ \hat{\mathbf{a}}) +(V_{i} - \tilde{\mu}_{i}) u_{n}(\mathbf{r}_{i})$, $\hat{\mathbf{a}}$ is the vector pointing to the nearest neighbors, and $\hat{\Delta} u_{n}(\mathbf{r}_{i}) = \Delta(\mathbf{r}_{i}) u_{n}(\mathbf{r}_{i})$. An analogous relation holds for the $v_{n}(\mathbf{r}_{i})$'s. Working at zero temperature $T=0$, we obtain the self-consistency equations
\begin{eqnarray}\label{SelfConsistency}
\Delta(\mathbf{r}_{i}) &=& U\sum_{n} u_{n}(\mathbf{r}_{i})v^{*}_{n}(\mathbf{r}_{i}),\\
\left< n_{i} \right> &=& 2 \sum_{n} |v_{n}(\mathbf{r}_{i})|^{2}.
\end{eqnarray}
 Starting with an ansatz for $\Delta({\mathbf{r}_{i}})$ and $n_{i}$, i.e., an ansatz for $\tilde{\mu}_{i}$ and $\Delta(\mathbf{r}_{i})$, we solve the BdG equations (\ref{BdGeqn}) on a honeycomb lattice with periodic boundary conditions. Doing so, we obtain the eigenenergies $\epsilon_{n}$ and the wave functions $u_{n}(\mathbf{r}_{i})$ and $v_{n}(\mathbf{r}_{i})$. We iterate this process until the solutions for the pairing amplitude and number of fermions satisfy the self-consistent equations (\ref{SelfConsistency}) at each lattice site to an accuracy of at least $5$ percent. The chemical potential $\mu$ is chosen such that the effective $\tilde{\mu}_{i}$ containing the Hartree shift keeps the average density of particles in the system $\left <n \right>=1$ up to a precision $\sqrt{V}$, where $V$ is the width of the disorder distribution. We average the results over 10-15 disorder realizations for each given $V$.

\section{PHASE DIAGRAM}
\label{sec:Clean}

\begin{figure}[t]
\includegraphics[width=\columnwidth]{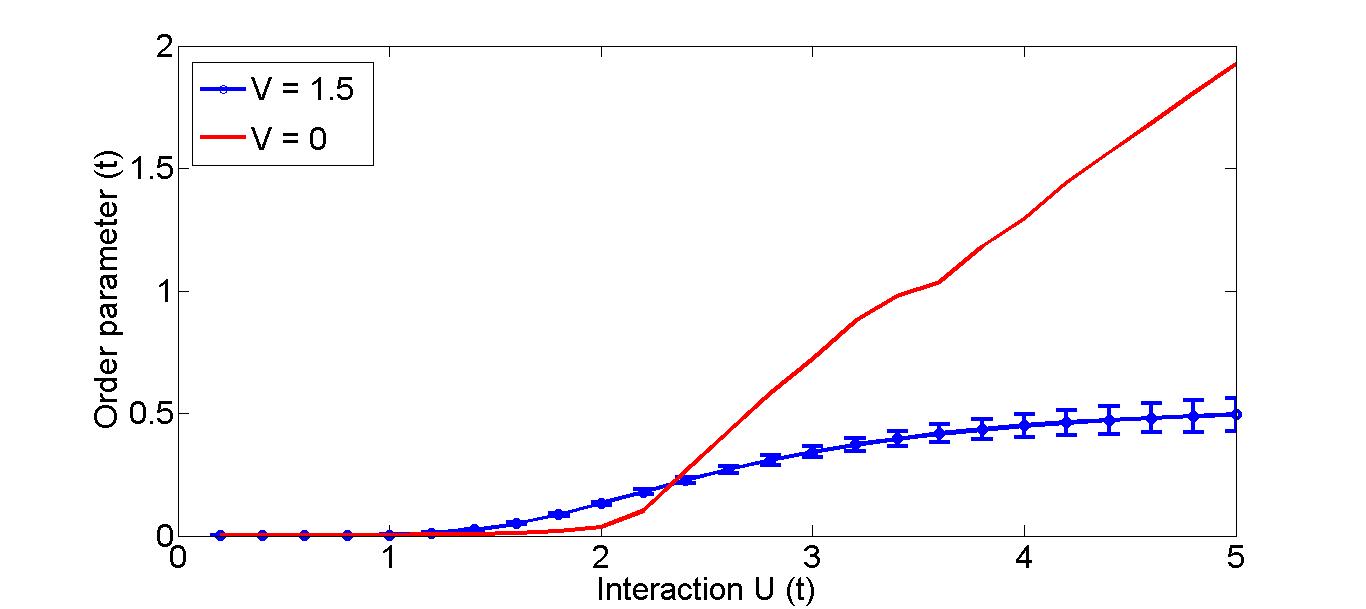}
\caption{Order parameter as a function of the attractive interaction strength $\Delta_\textrm{op}(U)$, computed for the clean system (red line) and for nonzero disorder strength $V=1.5$ (blue line). The results for the disordered system are averaged over 12 disorder realizations.}
\label{fig:turningon}
\end{figure}

In Fig.~\ref{fig:turningon} we plot the order parameter $\Delta_\textrm{op}$, defined as the spatial average of the local pairing amplitude $\Delta_\textrm{op} = \sum_{i} \Delta(\mathbf{r}_{i})$. \cite{ghosal_role_1998,ghosal_inhomogeneous_2001} We confirm the well-known result\cite{wilson1973,gross1974,kopnin2008,uchoa2009,zhao2006,roy2013} that in the absence of disorder, there exists a critical value $U_{c}$ for the attractive interactions that separates the Dirac semimetal phase from the superconducting one: in the former the order parameter is zero, whereas in the latter it is non-zero. We note that in the absence of disorder the superconductivity is spatially homogeneous.

In the presence of disorder, the Dirac semimetal phase is destroyed and the system becomes superconducting even below the clean attraction threshold $U_c\approx 1.8$ (Fig.~\ref{fig:turningon}). In principle, this should occur for arbitrarily weak disorder,\cite{nandkishore_superconductivity_2013} but this is not visible in our numerical simulations due to finite-size effects. Indeed, for sufficiently small $U$ the superconducting coherence length will be greater than the system sizes we are using. However, for $U$ close to but {\it less} than $U_c$, we observe that the disordered system is superconducting while the clean system is not, i.e., superconductivity is enhanced by disorder in this regime. Furthermore, we find that for $U>U_c$ disorder suppresses superconductivity.

\begin{figure}[t]
\centering
\includegraphics[width=\columnwidth]{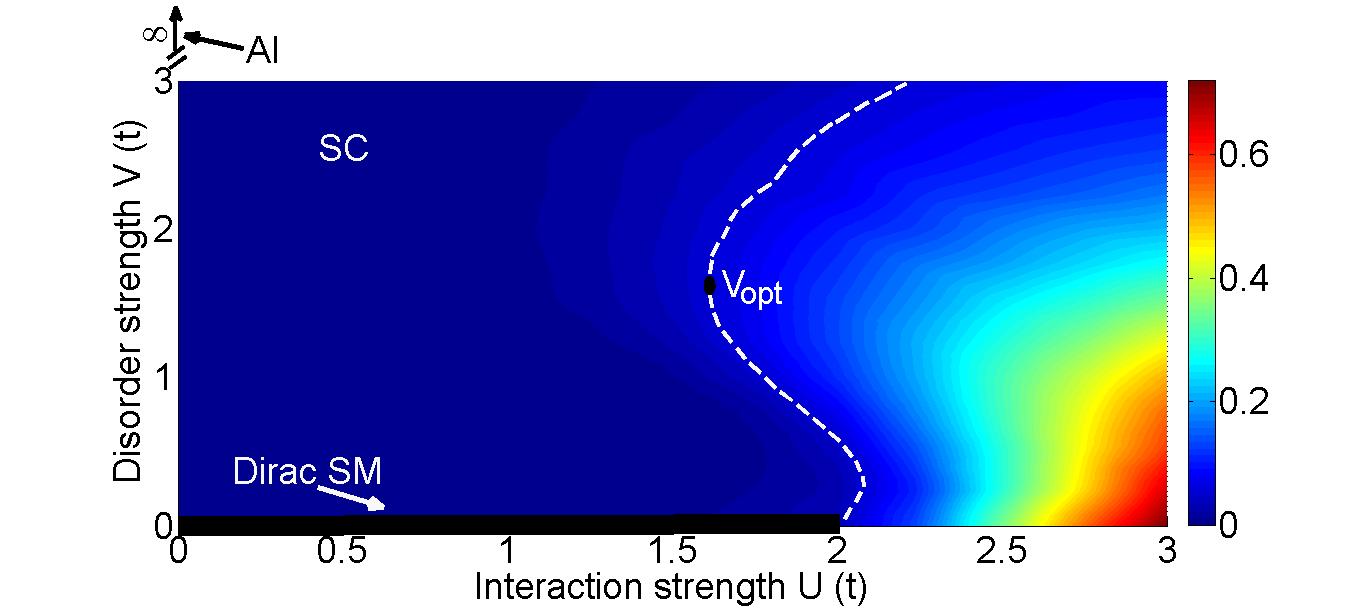}
\caption{Phase diagram ($\Delta_{\textrm{op}}$) in the $V$-$U$ plane: SC is the superconductor; Dirac SM is the Dirac semimetal; AI is the Anderson insulator; $V_{\textrm{opt}}$ is the optimal disorder. Below the critical coupling ($U<U_c \approx 1.8$), weak disorder enhances superconductivity, while strong disorder suppresses it. In this regime there is an optimal disorder $V_{\textrm{opt}} \approx 1.5$ strength that maximizes superconductivity. For stronger interactions $U>U_c$, disorder suppresses superconductivity.}
\label{fig:phasediagram}
\end{figure}

The full phase diagram in the $V$-$U$ plane (Fig.~\ref{fig:phasediagram}) encapsulates the interplay of attractive interactions, disorder, and superconductivity. As mentioned previously, for small attractive interactions ($U\lesssim 1.0$) the superconducting phase is not visible in the numerics due to finite-size effects. However, for a substantial regime near the clean attraction threshold $U\sim U_{c}$, we observe that weak disorder enhances superconductivity, but strong disorder suppresses it, such that there is an optimal disorder strength which maximizes superconductivity. We believe the suppression of superconductivity by strong disorder to be a signature of Anderson localization physics. Since a single Dirac fermion is protected against Anderson localization, we conjecture that the effect of disorder on superconductivity in the weak attraction regime will be monotonic for a single time-reversal invariant Dirac fermion. However, we are unable to test this numerically, since one cannot obtain a single Dirac fermion in a 2D lattice Hamiltonian without breaking parity and time-reversal symmetry.\cite{nielsen1981a,nielsen1981b,niemi1983,redlich1984} (One can design a time-reversal invariant bilayer lattice model with a pair of Dirac fermions such that for momenta close to the Dirac point, the Dirac fermions are localized on opposite surfaces and are approximately decoupled.\cite{marchand2012} However, the coupling between the two Dirac fermions increases as one moves away from the Dirac point, such that in the presence of strong interactions it is not clear that one can reliably mimick the superconductivity of a single Dirac fermion in such a model.) Meanwhile, for $U\gg U_{c}$ where the clean system was already in the superconducting phase, disorder always suppresses superconductivity.

\section{Inhomogeneity of the disordered superconductor}
\label{sec:Inhomogeneity}

\begin{figure}[t]
\centering
\includegraphics[width=\columnwidth]{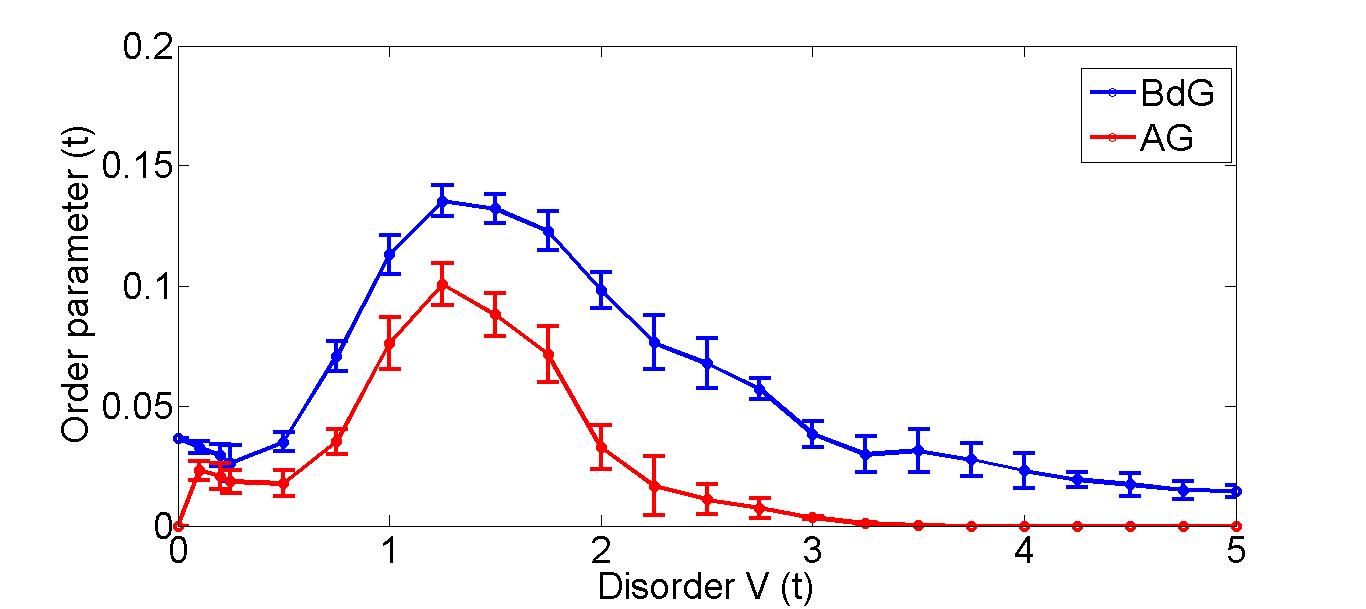}
\caption{Comparison of the uniform Abrikosov-Gorkov order parameter $\Delta_{AG}$ (red curve) and the spatially inhomogeneous BdG order parameter $\Delta_\textrm{op}$ (blue curve) as functions of the disorder strength $V$. Results are averaged over 12 disorder realizations for each disorder strength $V$, and are computed for $U \sim U_{c}$.}
\label{fig:AGBDG}
\end{figure}

In this section we show that superconductivity in the disordered $U<U_c$ regime is highly inhomogeneous, and should be thought of in terms of rare superconducting puddles that eventually establish global phase coherence.\cite{nandkishore_superconductivity_2013}

The Abrikosov-Gorkov (AG) theory\cite{Abrikosov-Gorkov} provides a framework for analyzing superconductivity in disordered systems, but assumes that the superconducting order parameter is translationally invariant. We compare the predictions of this theory with the results from an explicit solution of the BdG equations, and show that it dramatically underestimates the strength of superconductivity in the disordered system. We assume a spatially uniform order parameter $\Delta_\textrm{AG} = \frac{1}{N}\sum_{i}\Delta(\textbf{r}_{i}) = \Delta(\textbf{r}_{i})$ and roughly follow Anderson's treatment of dirty superconductors in the absence of magnetic impurities.\cite{Anderson} If we consider $w_{n}(\mathbf{r}_{i})$ to be the eigenfunctions of $\hat{H}_{K}$ with eigenvalues $\lambda_{n}$, we can set $u_{n}(\mathbf{r}_{i}) = u_{n} w_{n}(\mathbf{r}_{i})$ and $v_{n}(\mathbf{r}_{i}) = v_{n} w_{n}(\mathbf{r}_{i})$. The BdG equations immediately yield $\epsilon_{n}^{2} = \lambda_{n}^{2} + |\Delta_\textrm{AG}|^{2}$. Using the self-consistency and normalization conditions, we obtain
\begin{equation}
\Delta (\textbf{r}_{i}) = U \sum_{n} |w_{n}(\mathbf{r}_{i})|^{2} \frac{\Delta_\textrm{AG}}{2\sqrt{\Delta_\textrm{AG}^{2}+ \lambda_{n}^{2}}}
\end{equation}
Introducing the local density of states (LDOS) in the normal state $\rho(\mathbf{r}_{i},\omega) = \sum_{n}|w_{n}(\mathbf{r}_{i})|^{2} \delta(\omega-\lambda_{n})$, we get the following equation for $\Delta_\textrm{AG}$ which is similar to the gap equation for $s$-wave superconductivity in a clean metal,
\begin{equation}\label{AG}
1 = U\int d\omega\frac{\rho(\omega)}{2\sqrt{\omega^{2}+ \Delta_\textrm{AG}^2}},
\end{equation}
where $\rho(\omega)=\frac{1}{N}\sum_i\rho(\b{r}_i,\omega)$ is the total density of states (DOS) in the normal state. For various disorder strengths $V$, we have numerically computed the BdG order parameter $\Delta_\textrm{op}$ without assuming spatial homogeneity, and also the AG order parameter $\Delta_\textrm{AG}$ from Eq.~(\ref{AG}). As seen in Fig.~\ref{fig:AGBDG}, the uniform AG mean-field theory underestimates the strength of superconductivity in comparison to an approach that allows for spatial inhomogeneity.

\subsection{Superconducting islands}
\label{sec:SCislands}

\begin{figure}[t]
\centering
\subfigure[\ $U=5.0$, $V=0.25$, and $\Delta_\textrm{op}=1.84$]{\includegraphics[width=0.49\columnwidth]{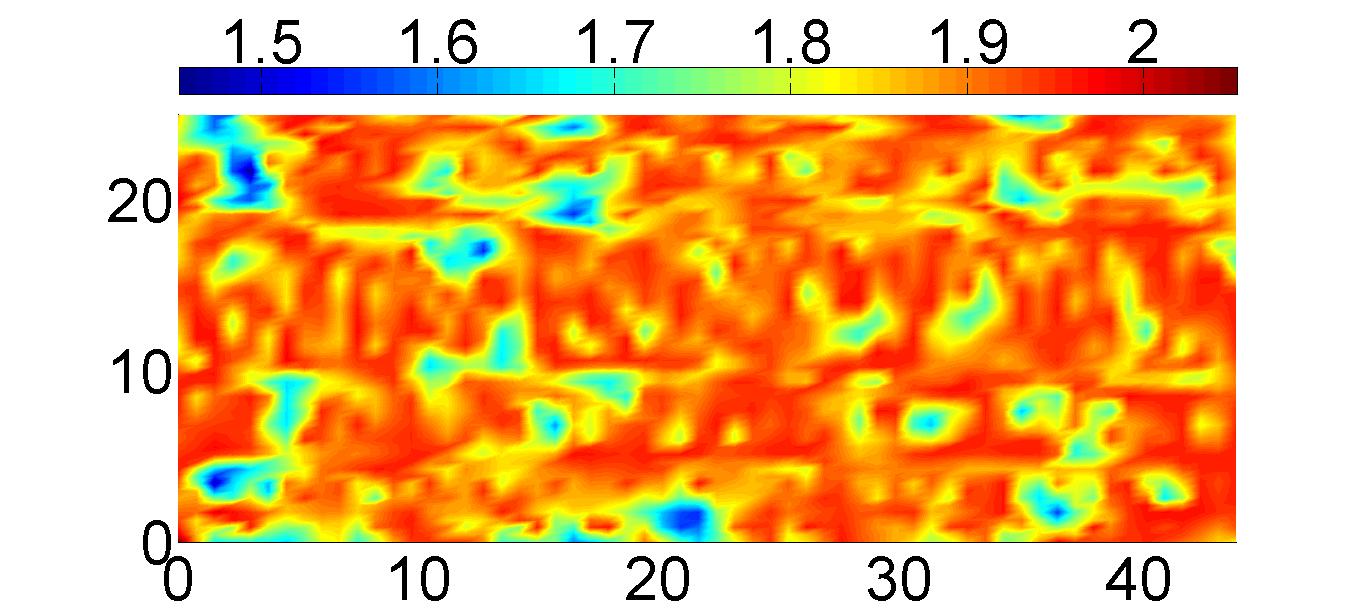}}
\subfigure[\ $U=5.0$, $V=5.0$, and $\Delta_\textrm{op} = 0.05$]{\includegraphics[width=0.49\columnwidth]{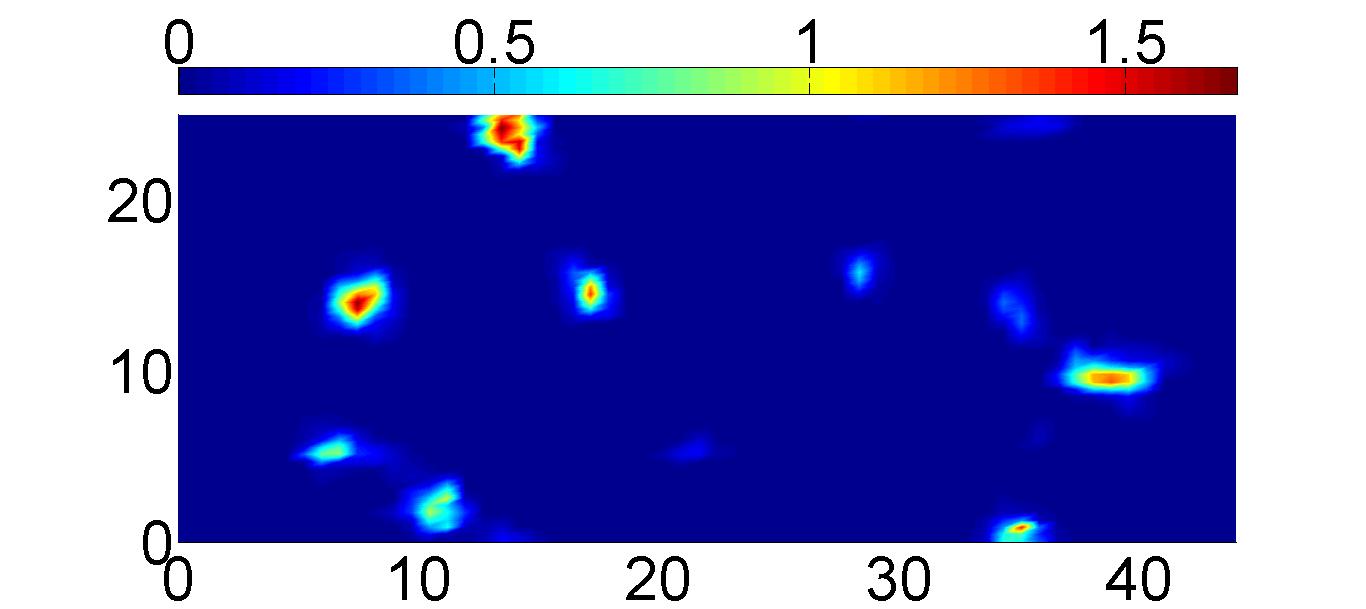}}
\subfigure[\ $U=0.8$, $V=1.0$, and $\Delta_\textrm{op} \sim 10^{-3}$]{\includegraphics[width=0.49\columnwidth]{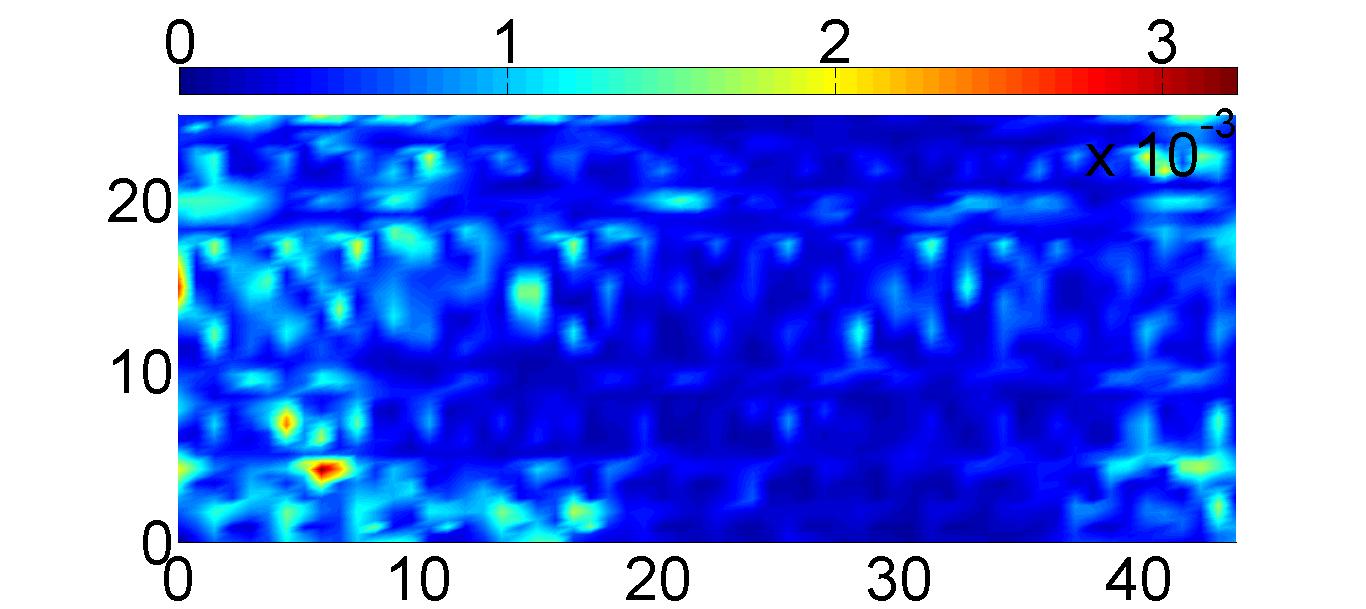}}
\subfigure[\ $U=0.8$, $V=5.0$, and $\Delta_\textrm{op} \sim 10^{-3} $]{\includegraphics[width=0.49\columnwidth]{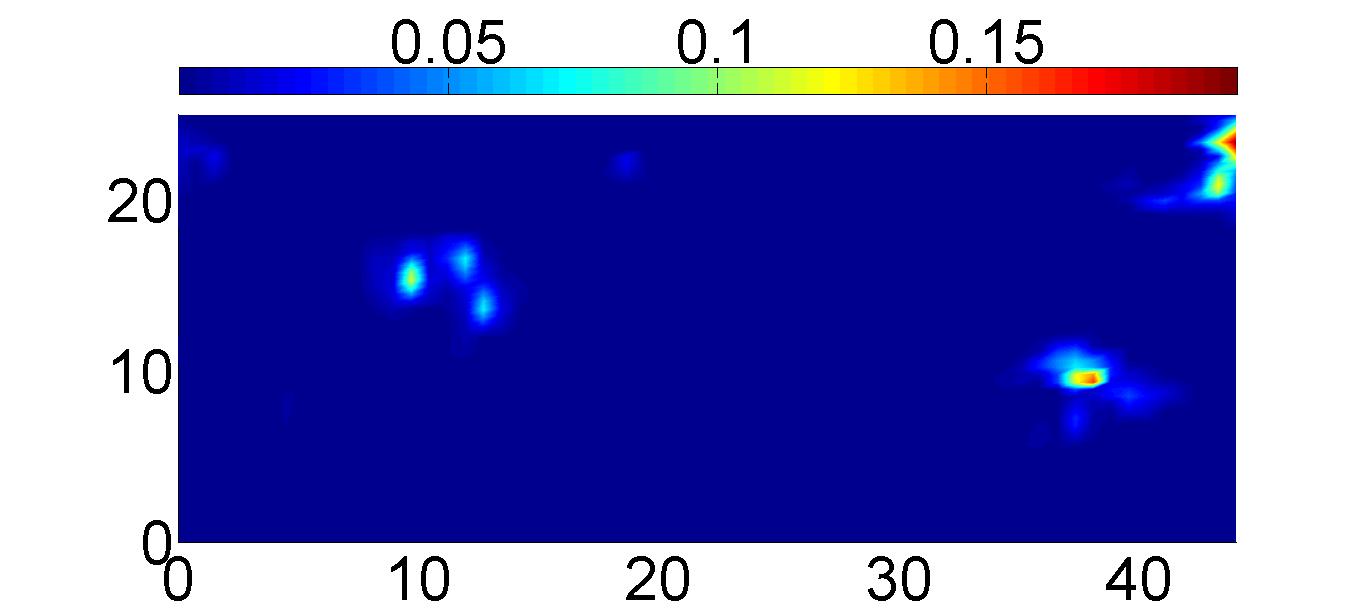}}
 \caption{Spatial distribution of the pairing amplitude in various regimes of interaction and disorder (blue: semimetal, red: superconductor). Strong disorder leads to the formation of superconducting islands, clusters of high pairing amplitude surrounded by a sea of small amplitudes.}
  \label{fig:pairing}
\end{figure}

\begin{figure}[t]
\centering
\includegraphics[width=\columnwidth]{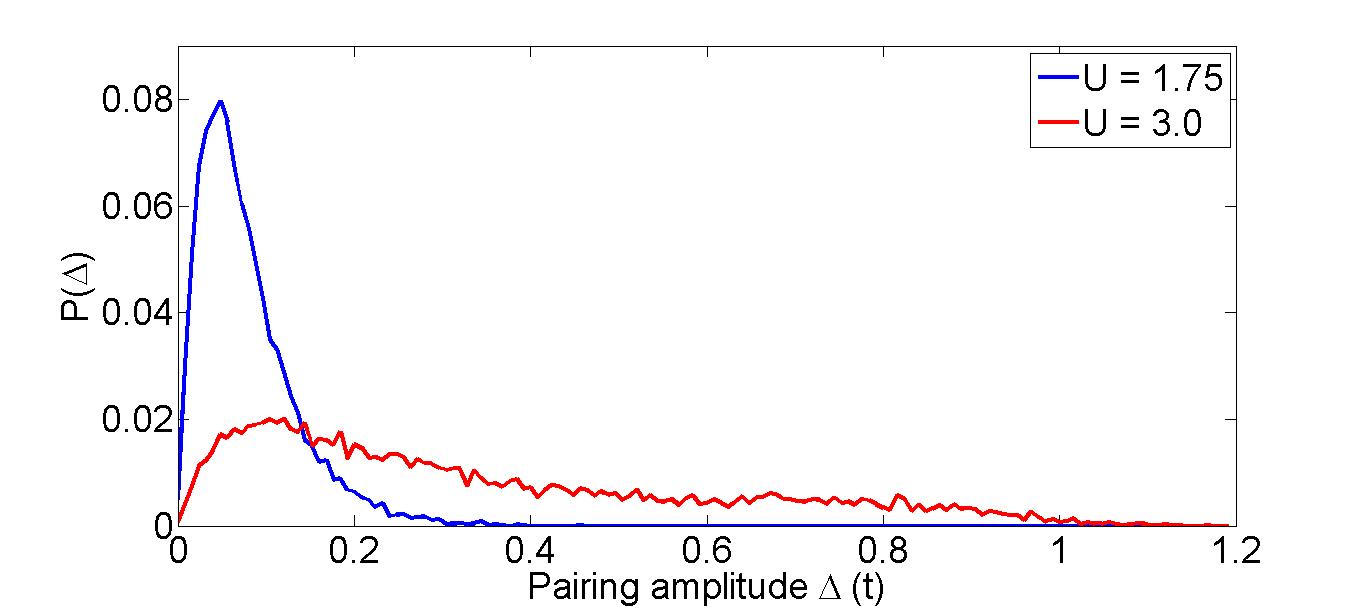}
\caption{Distribution of pairing amplitudes $P(\Delta)$ for $V=1.5$ (averaged over 12 disorder realizations) for $U=1.75<U_{c}$ (blue curve) and $U = 3.0>U_{c}$ (red curve).}
\label{fig:deltadistribution}
\end{figure}

To understand numerically how mesoscopic fluctuations and superconducting islands yield a more robust superconductivity than predicted by a spatially uniform pairing amplitude, we analyze the spatial distribution of $\Delta(\mathbf{r}_{i})$ in various regimes of attraction and disorder strengths. We determine the spatial distribution of pairing strengths well above $U_{c}$ [Fig.~\ref{fig:pairing}(a),(b)] and well below $U_c$ [Fig.~\ref{fig:pairing}(c),(d)]. Well above the critical point ($U = 5.0 >U_{c}$), the clean system is superconducting with a spatially uniform pairing amplitude. Weak disorder ($V = 0.25$) breaks the spatial uniformity of $\Delta(\mathbf{r}_{i})$ [Fig.~\ref{fig:pairing}(a)]. For a disorder strength that is comparable to the interaction strength ($V=U=5.0$), superconducting islands emerge, i.e., clusters of high pairing amplitude surrounded by a sea of small pairing amplitudes [Fig.~\ref{fig:pairing}(b)]. Meanwhile, well below the critical point ($U = 0.8 < U_{c}$), the clean system is not superconducting. In the presence of weak disorder ($V=1.0$) however, superconductivity emerges in a few rare regions [Fig.~\ref{fig:pairing}(c), red spots]. These superconducting islands become more clearly visible when the disorder strength is increased [Fig.~\ref{fig:pairing}(d)].

To study the behavior around $U_c$, we have looked at the pairing amplitude distribution\cite{ghosal_role_1998,ghosal_inhomogeneous_2001} $P(\Delta)$ for a given disorder strength $V=1.5$ (averaged over 10-15 realizations) and for two representative interaction strengths: $U_{1} = 1.7 <U_{c}$ and $U_{2} = 3.0 > U_{c}$ (Fig.~\ref{fig:deltadistribution}). As the interaction strength is increased from below to above $U_{c}$, the distribution $P(\Delta)$ becomes broader. For $U<U_c$ superconductivity is concentrated in a few regions, whereas for $U > U_c$ most of the sample is superconducting whereas a few regions are not.  

\subsection{Density of states and spectral gap}
\label{sec:DOS}

\begin{figure}[t]
\centering
\subfigure[\ $U=4.5 > U_{c}$]
{\includegraphics[width=0.49\columnwidth]{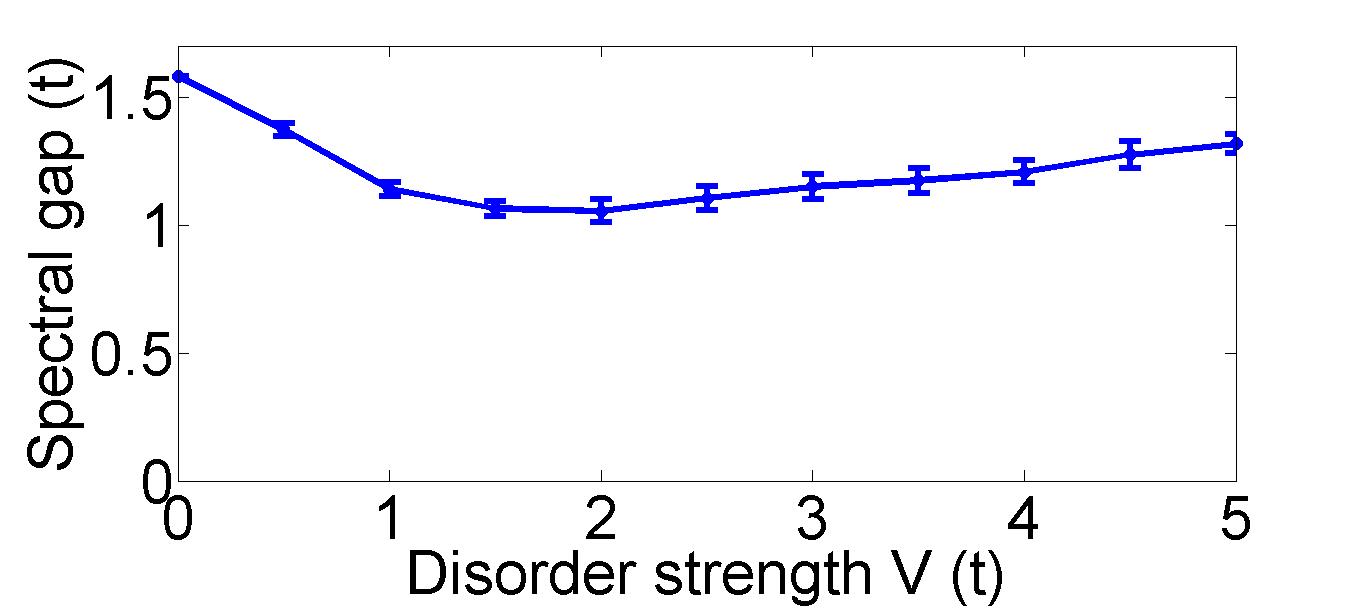}}
\subfigure[\ $U=1.0 < U_{c}$]
{\includegraphics[width=0.49\columnwidth]{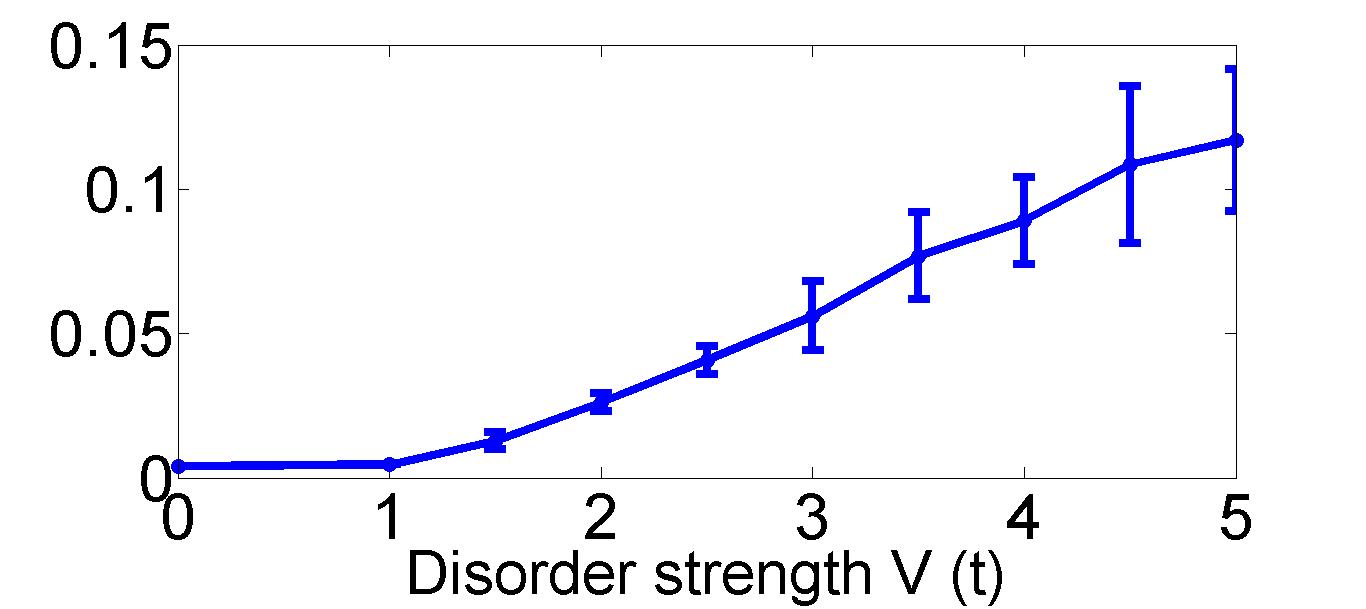}}

  \caption{Spectral gap $\epsilon_{\mathrm{gap}}$ in the DOS [Eq.~(\ref{DOS})] as a function of disorder strength for different interaction strengths $U$.}
  \label{fig:spectral}
\end{figure}

We now study the spatial distribution of the single-particle DOS defined as\citep{ghosal_inhomogeneous_2001}
\begin{equation}\label{DOS}
\rho(\omega) = \frac{1}{N} \sum_{n,\mathbf{r}_{i}}\left(|u_{n}(\mathbf{r}_{i})|^{2}\delta(\omega-\epsilon_{n}) + |v_{n}(\mathbf{r}_{i})|^{2}\delta(\omega+\epsilon_{n})\right),
\end{equation}
In the numerical calculations delta functions are replaced by a narrow Lorentzian lineshape.

Around the critical interaction strength ($U=1.8 \sim U_{c}$), the DOS (computed as the spatial average of the local density of states as seen in Eq.~[\ref{DOS}]) has different profiles depending on the disorder strength $V$ (Fig.~\ref{fig:LDOS}). For the clean system, the DOS has two well-resolved coherence peaks [Fig.~\ref{fig:LDOS}(a)], but it does not have a hard gap due to the smoothing function we have employed. For stronger disorder the DOS becomes ``smeared'' as higher energy states become available and it retains a gap around the zero energy point [Fig.~\ref{fig:LDOS}(b)]. In order to study the behavior of the gap $\epsilon_{\mathrm{gap}}$ in the DOS, we also look at the lowest eigenvalue of the BdG Hamiltonian matrix in Eq.~(\ref{BdGeqn}).

Well above the critical point ($U=4.5 >U_{c}$), the evolution of $\epsilon_{\mathrm{gap}}$ with disorder is non-monotonic [Fig.~\ref{fig:spectral} (a)], and has been explained by Ghosal \emph{et al.}\cite{ghosal_role_1998,ghosal_inhomogeneous_2001} Initially disorder suppresses superconductivity, which is reflected in a decrease of the spectral gap. However, a nonzero gap survives even for large disorder strengths because there are islands of superconductivity in the areas where $|V_{i}|$ is small and where particle-hole mixing occurs, whereas where $|V_{i}|$ is high the pairing amplitude vanishes (in the high mountains there are no electrons and in the deep valleys there are two). We have also found that the low-energy excitations lie entirely on the superconducting islands\citep{ghosal_inhomogeneous_2001} which explains the finite spectral gap at high disorder strengths. Below the critical point ($U=1.0< U_{c}$), the spectral gap is a monotonically increasing function of disorder [Fig.~\ref{fig:spectral}(b)]. This indicates that for subcritical couplings, disorder enhances pairing.

\subsection{Local density of states}
\label{sec:LDOS}

\begin{figure}[t]
\centering
\subfigure[\ $U=1.8$, $V=0$]
{\includegraphics[width=0.49\columnwidth]{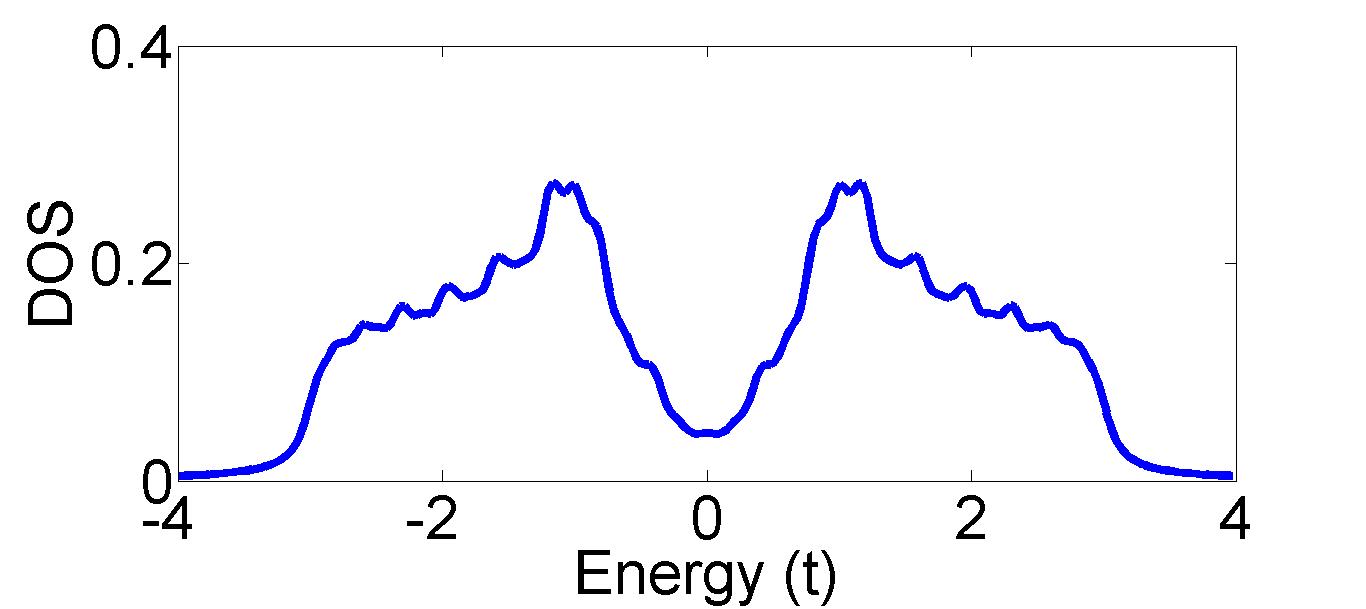}}
\subfigure[\ $U=1.8$, $V=1.5$]
{\includegraphics[width=0.49\columnwidth]{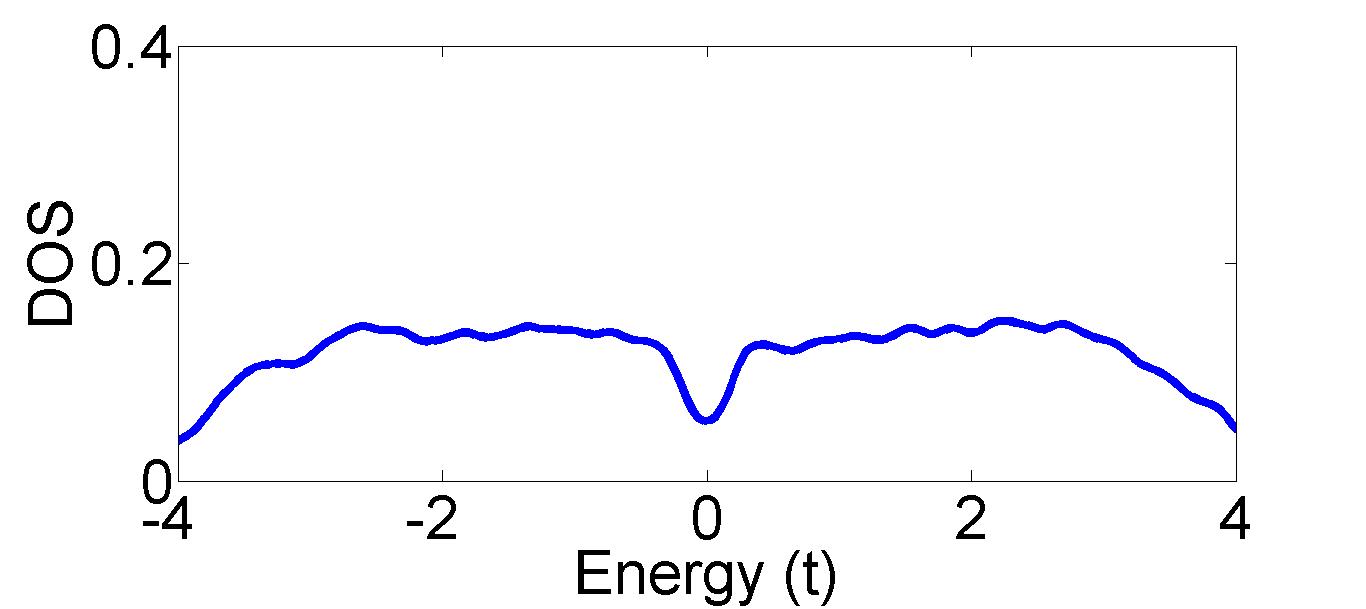}}
\subfigure[\ $U=1.8$, $V=0.5$, $\omega=1.0$]
{\includegraphics[width=0.49\columnwidth]{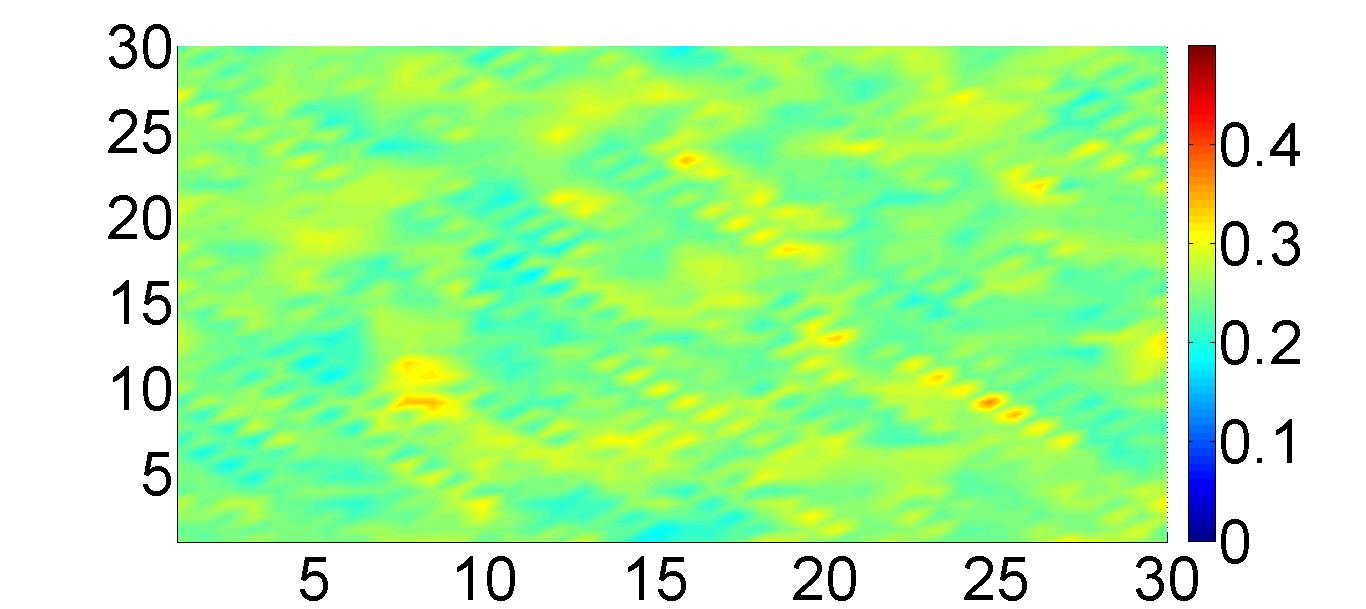}}
\subfigure[\ $U=1.8$, $V=1.5$, $\omega=1.0$]
{\includegraphics[width=0.49\columnwidth]{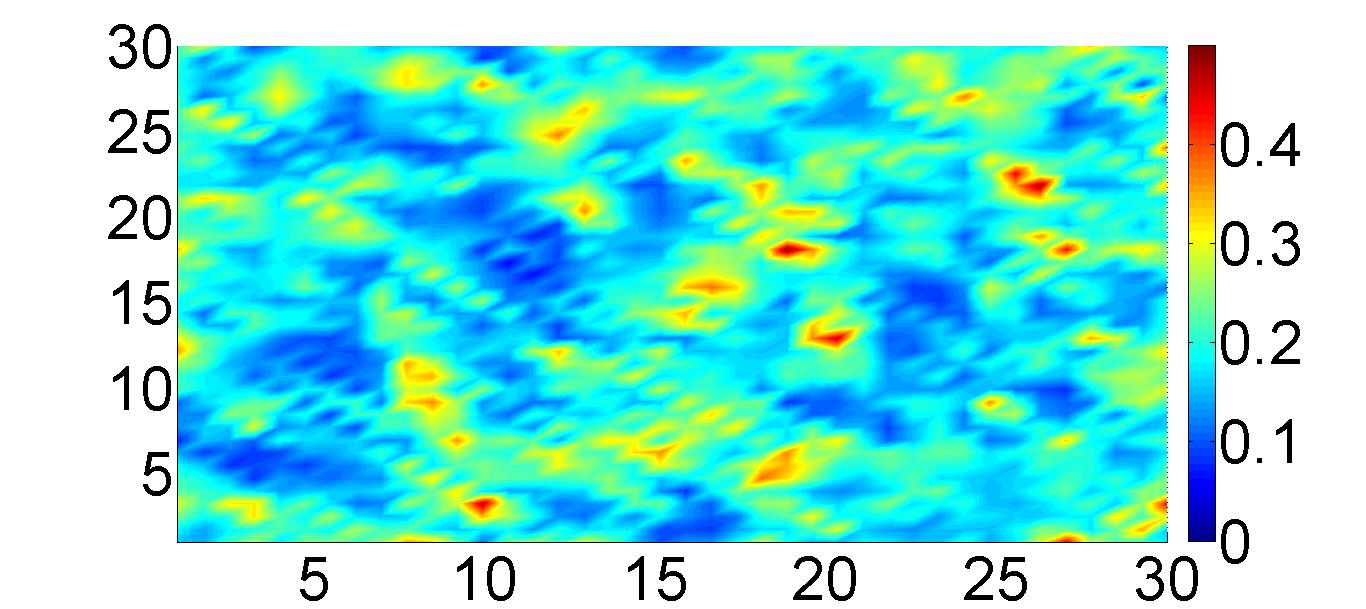}}
  \caption{The spatial average of the LDOS [(a),(b)] and the spatial plot of the LDOS at a fixed energy $\omega$ for the same disorder realization [(c),(d)].}
  \label{fig:LDOS}
\end{figure}

We also compute the local density of states (LDOS), defined as
\begin{equation}
\rho(\mathbf{r}_{i},\omega) = \sum_{n} \left(|u_{n}(\mathbf{r}_{i})|^{2}\delta(\omega-\epsilon_{n}) + |v_{n}(\mathbf{r}_{i})|^{2}\delta(\omega+\epsilon_{n})\right),
\end{equation}
for a given realization of disorder.

A way of studying the LDOS which is akin to STM experiments \cite{STMexperiments, gomes, hoffman} is to make spatial plots of the LDOS scanned at a fixed energy $\omega$ [Fig.~\ref{fig:LDOS}(c),(d)]. In the weak disorder regime, the LDOS is roughly spatially uniform. In this regime the enhancement of superconductivity is too weak to be visible. In the strong disorder case where the enhancement of superconductivity is more visible, we observe a spatially inhomogeneous LDOS that we interpret as a signature of the formation of superconducting islands. The observation of superconducting islands in the LDOS confirms our expectations that disorder-enhanced superconductivity in the subcritical attraction regime should be highly spatially inhomogeneous.

\section{PROXIMITY-INDUCED SUPERCONDUCTIVITY}
\label{sec:Proximity}

\begin{figure}[t]
\centering
\includegraphics[width=\columnwidth]{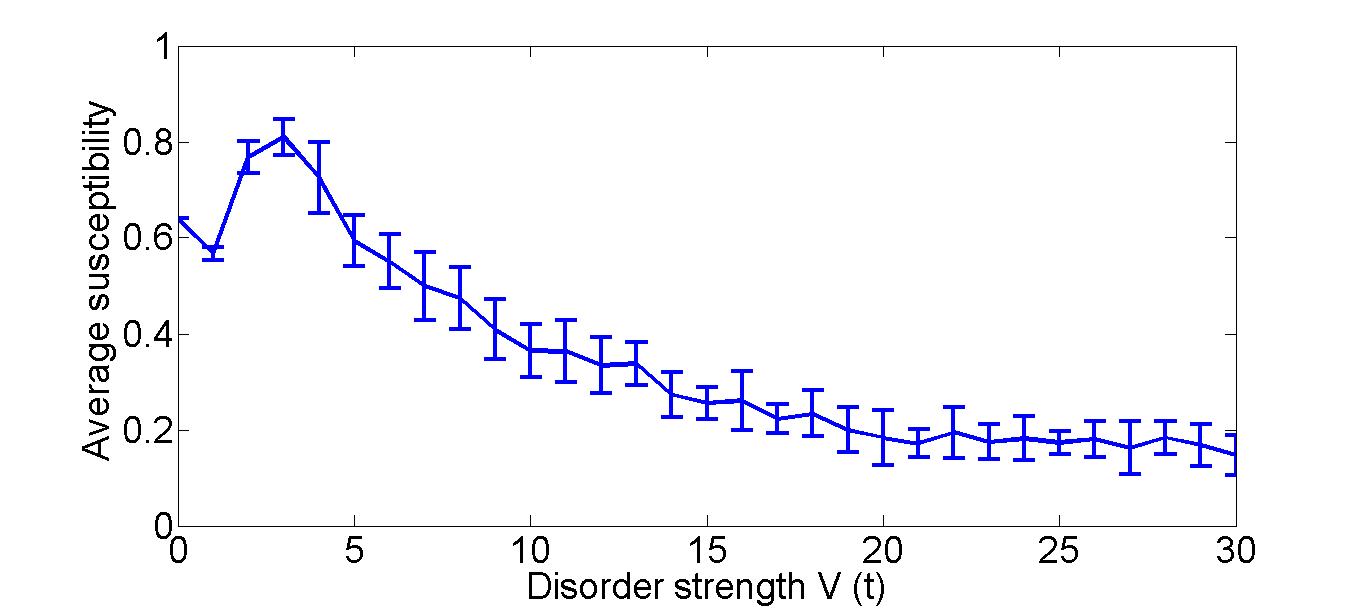}
\caption{Average superfluid susceptibility $\bar{\chi}$ as a function of disorder strength $V$ for proximity-induced superconductivity in a dirty graphene sheet.}
\label{fig:spatialsusceptibility}
\end{figure}

An important experimental application of the enhancement of superconductivity by disorder is the superconducting proximity effect. In particular, one would expect an enhancement of the superfluid susceptibility in a sheet of dirty graphene proximate to a superconductor. To model this, we consider the Hamiltonian for a disordered graphene lattice with no attractive interactions, i.e., Eq.~(\ref{Hamiltonian}) with $U=0$,
\begin{equation}
H = -t\sum_{\left<ij\right>, \sigma} (c_{i\sigma}^{\dagger} c_{j\sigma} + \mathrm{H.c.}) + \sum_{i,\sigma}V_{i}n_{i\sigma}.
\end{equation}
Pairing occurs via the tunneling of Cooper pairs from the superconductor into the dirty graphene layer, which is modeled by an external, real, uniform pairing amplitude $\Delta(\mathbf{r}_{i}) = \Delta$. The full Hamiltonian becomes
\begin{align}\label{ProximityHamiltonian}
H_{\mathrm{proximity}}& = -t\sum_{\left<ij\right>, \sigma} (c_{i\sigma}^{\dagger} c_{j\sigma} + \mathrm{H.c.}) + \sum_{i,\sigma}V_{i}n_{i\sigma}\nonumber\\
& + \sum_{i}(\Delta c_{i\uparrow}^{\dagger}c_{i\downarrow}^{\dagger} + \mathrm{H.c.}).
\end{align}
 We define the local superfluid susceptibility as
\begin{equation}
\chi(\mathbf{r}_{i}) = \left .\frac{\partial}{\partial \Delta} \left|\left< c_{i\uparrow}^{\dagger}c_{i\downarrow}^{\dagger} \right>\right| \right|_{\Delta=0}.
\end{equation}
By computing the average local susceptibility $\bar{\chi} = \frac{1}{N}\sum_{i} \chi(\mathbf{r}_{i})$ as a function of disorder strength, we observe that there is an optimal regime (roughly $V\in[1,5]$) for which the susceptibility is enhanced compared to the clean case (Fig.~\ref{fig:spatialsusceptibility}). However, for very high disorder strengths ($V\gg5.0$), we recover the signature of an Anderson insulator \citep{Ma_Lee}, $\bar{\chi} \rightarrow 0$. This confirms that for subcritical interactions $U<U_c$, weak disorder enhances superconductivity while strong disorder suppresses it. Moreover, it demonstrates that this effect applies not only to intrinsic superconductivity, but also to proximity-induced superconductivity.

\section{Conclusions}
\label{sec:Conclusions}

We have demonstrated by a self-consistent numerical solution of the BdG equations on a disordered graphene lattice that for weak attractive interactions $U<U_c$, weak disorder enhances superconductivity. Thus, a disordered system can be superconducting even when a clean system is semimetallic. The effect is non-monotonic in the disorder strength, with strong disorder suppressing superconductivity, such that there is an optimal disorder strength that maximizes superconductivity. Moreover, superconductivity in this regime is spatially inhomogeneous, consisting of superconducitng islands immersed in a semimetallic sea. We have produced plots of the typical LDOS which may be directly compared to STM experiments in this regime. We have also shown that these effects apply to proximity-induced superconductivity as well as to intrinsic superconductivity. Meanwhile, for strong attractive interactions $U > U_c$, disorder suppresses superconductivity: this is the usual behavior, recovered here in the strong interaction regime. We anticipate that these results will be of relevance for ongoing experiments aiming to realize proximity-induced superconductivity in dirty graphene \cite{Heersche,han_collapse_2014}.

\acknowledgements

We are grateful to D. A. Huse and E. H. da Silva Neto for insightful discussions. This work was supported by the Simons Foundation (JM), a PCTS fellowship (RN) and by NSF Grant Numbers DMR 1006608 and 1311781 (SLS)

\bibliography{graphenepaper}

\end{document}